\documentclass{article}
\input{tcilatex}
\begin{document}

\title{A different perspective on the problem of time in quantum gravity\ \ }
\author{ \ \ M. Bauer$^{1\ast }$, C.A. Aguill\'{o}n$^{2}$ and G. Garc\'{\i}a$%
^{2}$ \\
%EndAName
1. Instituto de F\'{\i}sica; 2. Instituto de Ciencias Nucleares\\
Universidad Nacional Aut\'{o}noma de M\'{e}xico, CDMX, MEXICO\\
*bauer@fisica.unam.mx\ }
\maketitle

\begin{abstract}
The perspective is advanced that the time parameter in quantum mechanics
corresponds to the time coordinate in a Minkowski flat spacetime local
approximation to the actual dynamical curved spacetime of General
Relativity, rather than to an external Newtonian reference frame. There is
no incompatibility, as generally assumed in the extensively discussed
"problem of time" in Quantum Gravity.
\end{abstract}

\bigskip

\section{Introduction}

The ``problem of time" (PoT) in the quantization of General Relativity (GR)%
\cite%
{Anderson,Anderson1,Kuchar,Kiefer1,Ashtekar,Butterfield,Kiefer,Kiefer2,Isham1,Isham}
arises in part from considering that time in the Time Dependent Schr\"{o}%
dinger Equation (TDSE) is a parameter (an element of an external frame of
reference as the absolute space and absolute time introduced by Newton),
whereas in General Relativity (GR) time acquires a dynamical character due
to the curvature of spacetime that reflects the presence of mass.\ It is
thus stated that: ``On the other hand, one faces the problem of time.
Whereas time in quantum theory is absolute (the parameter in Schr\"{o}dinger
equation has been inherited from Newtonian mechanics), time as part of the
space-time obeying Einstein's equations is dynamical. A more fundamental
theory is therefore needed to gain a coherent concept of time" \cite{Kiefer1}%
; and ``The greater part of the PoT occurs because the ``time" of GR and the
``time" of Quantum Theory are mutually incompatible notions" \cite{Anderson}%
. This view is furthermore supported by Pauli's objection to the existence
of a time operator in Quantum Mechanics (QM) \cite{Pauli}.

\section{A different perspective}

Based on recent theoretical and experimental developments, this outlook of
the problem of time is modified if one takes into account the following:

a) The relativistic quantum equation as formulated by Dirac satisfies
Lorentz invariance. This is achieved by integrating the time parameter of
the time dependent Schr\"{o}dinger equation (TDSE) with the three space
coordinates\ into a four dimensional spacetime Minkowski frame of reference%
\cite{Thaller,Greiner}. As such, neither the time nor the space coordinates
are represented by operators, as postulated in quantum mechanics (QM) to be
associated to the observable properties of the system under study; nor time
can be considered to be canonically conjugate to the Hamiltonian \cite%
{Hilgevoord}.

b) Closed systems (foremost the Universe, although finite macroscopic and
atomic systems have been so considered for all practical purposes in the
formulations (non relativistic first and then relativistic) of classical and
quantum mechanics. Being closed, they are static, i.e., they do not evolve.
Therefor in QM the basic equation is the time independent Schr\"{o}dinger
equation (TISE).

c) Time in the TDSE is the laboratory time transferred by the entanglement
of the microscopic system with its macroscopic environment where clocks are
found; thus\ the TDSE is a classical quantum equation where $t$ is part of a
dynamical reference frame in the curved spacetime \cite%
{Briggs1,Briggs2,Moreva}. A similar situation is expected to follow in the
case of the time independent Wheeler-deWitt equation (WdW) in the canonical
quantization of GR \cite{Briggs1}.

d) QM and GR are at present \ assumed to be universally valid in the
cosmological development of the Universe, from the Big Bang to the
progressive expansion and cooling, the appearence of fundamental particles,
the aggregation into atoms and, as a consequence of decoherence, into the
massive components of the present world and the remnant cosmic background
radiation (CMBR), and possibly also dark matter and dark energy, all
inmersed in a GR curved spacetime \cite{Kiefer2}. Although this is still an
open problem, ``Local agreement with SR (Special Relativity) is also
required. A natural hypothesis here is Einstein's that SR inertial frames
are global idealizations of GR's local inertial frames that are attached to
freely falling particles. Furthemore, in parallel with the developmeent of
SR, Einstein retained a notion of metric $g_{\mu \nu }$ on spacetime to
account for observers in spacetime having the ability to measure lengths and
times if equipped with standard rods and clocks, encode the distinction
between time and space, as $g_{\mu \nu }$ reduces locally to GR's $\eta
_{\mu \nu }$ everywhere the other laws of Physics take their SR form"\cite%
{Anderson}. And also: ``Any acceptable quantum gravity theory must allow us
to recover the classical spacetime in the appropiate limit. Moreover, the
spacetime geometrical notions should be intrisically tied to the behavior of
matter that probes them" \cite{Bonde}.

e) A composite closed system ( micoscopic system plus macroscopic
environment -the laboratory-) is found in a curved spacetime representing
the pervasive gravitational interaction. However, the minimal (less than
0.02\%) correction to the hydrogen spectrum that arises from a Dirac
equation extended to curved spacetime suggests that the laboratory is
subject to a very weak curvature \cite{Barros}, so the Minkowski flat
spacetime of SR is a good local approximation to the GR curved spacetime.

Furthermore, Dirac's formulation of Relativistic Quantum Mechanics (RQM)
allows the introduction of a self-adjoint "time" operator for the
microscopic system $T=\mathbf{\alpha .r}/c+\beta \tau _{0}$, in analogy to
the Hamiltonian $H=c\mathbf{\alpha .p}+\beta m_{0}c^{2},$ where $\alpha
=(\alpha _{x,}$\ $\alpha _{y,}\alpha _{z})$\ and $\beta $\ \ are the Dirac
matrices. It represents in principle an additional observable. This operator
generates a unitary transformation that shifts momentum - whose spectrum is
continuous and unbounded -, and ensuingly the energy in both positive and
negative energy branches, thus circumventing\ Pauli's objection. It also
provides a time energy uncertainty relation directly related to the space
momentum uncertainty, as envisionned originally by Bohr in the uncertainty
of the time of passage of a wave packet at a certain point; and a formal
basis for de Broglie's daring assumption of associating a wave of frequency $%
\upsilon =mc^{2}/h$ to a particle of mass $m$ \cite{Bauer1,Bauer2,Bauer3,
Lan}.

\section{Conclusion}

In view of the above one can conclude that in relativistic classical and
quantum mechanics, time is part of a Minkowski reference frame that locally
approximates well the actual GR dynamic curved spacetime where the
laboratory is locates. It is not Newtonian. Its origin is dynamical\footnote{%
Another source of confusion arises because ``dynamical" is attached to two
different aspects. One is the variation associated with \ the impact of
matter on the spacetime reference frame. The other is the dependence on the
parameter $t$ of the observables of the system, that in QM is made explicit
in the Heisenberg picture.}. \textit{There is therefore no incompatibility}.
To quote Einstein: ``Newton, forgive me; you found the only way which, in
your age, was just about possible for a man of highest thought and creative
power. The concepts, which you created, are even today still guiding our
thinking in physics, although we now know that they will have to be replaced
by others farther removed from the sphere of immediate experience, if we aim
at a profounder understanding of relationships" \cite{Einstein}. This
apology and recognition should be extended to the creators of the Hamilton
Jacobi formulation of classical mechanics and to the creators of quantum
mechanics.

To be pointed out is that not every link of the present perspective has been
fully developed at present. The quantization of GR is still an open question
in many aspects \cite{Anderson,Bonde,Smolin}. However, this point of view
removes the question of whether time is to be identified before or after
quantization, in favour of a timeless interpretation of quantum gravity \cite%
{Isham}, where time would emerge as the observable that conditions all the
others, as proposed by Page and Wootters \cite{Page}. The proposed
self-adjoint time operator that complements the Dirac formulation of RQM may
play the role of that observable \cite{Bauer4}, in response to the objection
of Unruh and Wald \cite{Unruh}. And perhaps it may also help in removing the
ambiguity with respect to time in the foliation of spacetime in the
canonical approach to Quantum Gravity.\ 

.

\end{document}